\documentclass[pra,showpacs,amsmath,amssymb,twocolumn]{revtex4}

\usepackage{bm}

\begin{document}

\newcommand{\tr}{\mathop{\mathrm{tr}}\nolimits}

\preprint{}
\title{ Restriction on relaxation times \\
 derived from the Lindblad-type master equations for $2$-level systems}
\author{Gen Kimura}
\email{gen@hep.phys.waseda.ac.jp}
\affiliation{Department of Physics, Waseda University, Tokyo 169--8555, Japan}
\begin{abstract}
We discuss a restriction on relaxation times derived from the Lindblad-type master equations for $2$-level systems and show that none of the inverse relaxation times can be greater than the sum of the others. The relation is experimentally proved or disproved and can be considered to be a measure for or against the applicability of the Lindblad-type master equations and therefore of the so-called completely positive condition.
\end{abstract}
\maketitle


There are still many important problems on the foundation of Quantum Mechanics left open. One of them is how to understand non-unitary time evolutions, such as the transition to thermal equilibrium state or decoherence process in quantum measurement. For example, the reduced dynamics that takes the concept of an environment into account may give some answer to this problem. The idea is as follows: Consider an environment $E$ which interacts with the system of interest $S$. Quantum mechanics is applied to the total system $S\otimes E$ so that the total system evolves unitarily. However, if we are only interested in the dynamics of the system $S$, we are allowed to take a trace over the environment $E$, and the time evolution of the system $S$ turns out to be generally non-unitary \cite{ref:Gardiner,ref:Davies,ref:Alicki}.

A mathematical character for the reduced dynamics (derived from a system and its environment initially uncorrelated) is that the time evolution is represented by a completely positive map \cite{note:BH} at any time \cite{ref:Kraus}. Lindblad \cite{ref:Lindblad}, Gorini, Kossakowski, and Sudarshan \cite{ref:Gorini} derived the general form of a Markovian master equation \cite{ref:DS} that takes the complete positivity into account:
\begin{subequations}\label{eqn:Lindblad MEq}
\begin{equation}
\frac{d}{dt} \rho(t) = (\hat{{\mathcal H}} + \hat{{\mathcal D}}) \rho(t),
\end{equation}
\begin{eqnarray}\label{eqn:Lindblad Gen}
{\hat{\mathcal H}}\rho &=& -i \left[H, \rho \right],  \nonumber \\{\hat{\mathcal D}}\rho &=&  \frac{1}{2}\sum_{i,j=1}^{N^2-1}C_{ij}\left\{ \left[F_i,\rho F^{\dagger}_j \right] + \left[F_i\rho, F^{\dagger}_j \right] \right\},
\end{eqnarray}
\end{subequations}
where $\rho(t)$ is the state (density matrix) of the system $S$, $H$ a self-adjoint operator with $\tr H =0$, $F_i\ (i=1\sim N^2-1)$ some operators with $\tr F_i = 0$ and $\tr F^{\dagger}_iF_j = \delta_{ij}$ and $[C_{ij}]$ some positive matrix \cite{not:FinDim}. Notice that this positivity is direct requirement of the completely positive condition \cite{ref:Gorini}. The master equation (\ref{eqn:Lindblad MEq}) is often called the Lindblad-type master equation (or the master equation of the Kossakowski-generator, etc.) and has been widely used such as in quantum optics (atom and spin relaxation) \cite{ref:Alicki,ref:Gardiner}, quantum computation and quantum information \cite{ref:Ohya}, quantum measurement \cite{ref:Kraus}, quantum state diffusion \cite{ref:Percival}, dissipative quantum chaos \cite{ref:Braun} and particle physics \cite{ref:Benatti}. In some cases, a Lindblad-type master equation is used as a starting point, and its use usually seems to have been taken for granted. In this circumstance, the existence of a practical criterion, capable for experimental investigations to test the validity of any Lindblad-type master equations, would be quite helpful. In this letter, we show that there is a universal law among relaxation times $T_i \ (i=1,2,3)$ in the Lindblad-type master equations for $2$-level systems, the fact of which generalizes the result derived by Gorini {\it et~al.} \cite{ref:Gorini}. This will serve as the above criterion.

We begin by reviewing the result derived by Gorini {\it et~al.}, giving clear looks at its background. In Ref.~\cite{ref:Gorini}, they discussed the dynamical properties of the Lindblad-type master equations for $2$-level systems ($N=2$) and showed that under the condition
\begin{equation}\label{eqn:Condition}
\left[{\hat{\mathcal H}},{\hat{\mathcal D}} \right] = 0,
\end{equation}
the following relations among the inverse relaxation times ${\mit\Gamma}_i = 1/T_i\ (i=1\sim 3)$ hold:
\begin{equation}\label{eqn:RestrictionAboutGamma}
{\mit\Gamma}_i + {\mit\Gamma}_j \ge {\mit\Gamma}_k \ge 0, \quad (i,j,k) :\  \mathrm{a\ permutation \ of\ (1,2,3)}.
\end{equation}
A familiar example which satisfies condition (\ref{eqn:Condition}) is a master equation equivalent to the following Bloch equation \cite{ref:Bloch,ref:2GT>=GL}:
\begin{subequations}\label{eqn:BlochEq1}
\begin{equation}
\frac{d}{dt}\bm{M}(t) = - A \bm{M}(t) + \bm{b},
\end{equation}
\begin{equation}
A =  \left(
\begin{array}{ccc}
1/T_T & -\Omega & 0 \\
\Omega & 1/T_T& 0 \\
0&0&1/T_L
\end{array}
\right), \ \bm{b} = \left(
\begin{array}{ccc}
0 \\
0 \\
k
\end{array}
\right),
\end{equation}
\end{subequations}
where $\bm{M}(t) = {}^{t}(M_x(t),M_y(t),M_z(t)), \ M_i(t)=\tr \rho(t) \sigma_i/2 \ (i=x,y,z)$ are polarization components, $\sigma_i$'s being the Pauli matrices \cite{note:BlochEq}. The inverse relaxation times ${\mit\Gamma}_i\ (i=1,2,3)$ are defined as the real parts of eigenvalues of the matrix $A$ in (\ref{eqn:BlochEq1}), so that the relaxation times $T_i \equiv 1/{\mit\Gamma}_i$ give the time scales of an exponential decay. Since the eigenvalues of $A$ in (\ref{eqn:BlochEq1}) are $1/T_L$ and $1/T_T \pm i \Omega$, we have two relaxation times, $T_L$ (longitudinal relaxation time) and $T_T$ (transverse relaxation time). Condition (\ref{eqn:Condition}) can be easily shown to be satisfied, i.e., the equivalent master equation satisfies condition (\ref{eqn:Condition}), thus relations (\ref{eqn:RestrictionAboutGamma}) hold from the result derived by Gorini {\it et~al.} Notice that, in this case, they are equivalent to the famous relation:
\begin{equation}\label{eqn:RestrictionAboutTLandTT}
2T_L \ge T_T \ge 0,
\end{equation}
usually observed experimentally \cite{ref:Alicki,ref:2GT>=GL}.


Recently, another interest has been raised in the ``strong'' interaction between system and its environment \cite{ref:Gardiner2,ref:Gen1}. In such cases, condition (\ref{eqn:Condition}) does not generally hold and it seems that relations (\ref{eqn:RestrictionAboutGamma}) are not satisfied either. However, master equations that are of the Lindblad-type and have the inverse relaxation times that satisfy (\ref{eqn:RestrictionAboutGamma}) have been found \cite{ref:Gen1}, despite the fact that condition (\ref{eqn:Condition}) is not satisfied. This means that the condition is not indispensable for relations (\ref{eqn:RestrictionAboutGamma}) to hold and implies that the latter hold more widely in the Lindblad-type master equations. Then a natural question is how wide the validity of relations (\ref{eqn:RestrictionAboutGamma}) is. We shall show below that they are valid in any Lindblad-type master equations for $2$-level systems.

For a clear understanding, it would be convenient to follow the argument given by Gorini {\it et~al.} under condition (\ref{eqn:Condition}) \cite{ref:Gorini} and then to generalize it. First, we put $F_i = \sigma_i/2$ \cite{note:Normalization} and the positive matrix $[C_{ij}]$, written explicitly as
\begin{equation}\label{eqn:[C]}
[C_{ij}] = \left(
 \begin{array}{ccc}
 \gamma_2+\gamma_3 -\gamma_1 & -ia_3 & ia_2 \\
 ia_3 & \gamma_3 + \gamma_1 - \gamma_2 & -ia_1 \\
 -ia_2 & ia_1 & \gamma_1+ \gamma_2 -\gamma_3
\end{array}\right),
\end{equation}
where $ a_i$'s and $\gamma_i$'s are real parameters \cite{note:Axis}. We can expand $H$ in terms of $F_i$, $H = \sum_{i=1}^3 h_i F_i$, in which $h_i$'s are real parameters. Then the polarization components $M_i(t) = \tr \rho(t) F_i$ are easily shown to satisfy the generalized Bloch equation \cite{ref:Gorini,ref:DS2},
\begin{subequations}\label{eqn:MatrixBlochEq}
\begin{equation}
\frac{d}{dt} \bm{M}(t) = - A \bm{M}(t) + \bm{b},
\end{equation}
\begin{equation}
A=\left(
\begin{array}{ccc}
\medskip
 \gamma_1 &  h_3 & -h_2 \\
\medskip
-h_3 & \gamma_2 & h_1 \\
h_2 & -h_1 & \gamma_3
\end{array}
\right), \ {\bm b} = \frac{1}{2}\left(\begin{array}{ccc} a_1 \\ a_2 \\ a_3\end{array} \right),
\end{equation}
\end{subequations}
which is equivalent to the Lindblad-type master equation (\ref{eqn:Lindblad MEq}) for 2-level systems. Since the completely positive condition requires the matrix $[C_{ij}]$ (\ref{eqn:[C]}) to be positive, the diagonal matrix elements have to be positive, from which the positivity of $\gamma_i$'s themselves also follows \cite{note:PositivityOfgamma}, i.e., it holds that
\begin{equation}\label{eqn:RestrictionAboutgamma}
\gamma_i + \gamma_j \ge \gamma_k \ge 0, \quad (i,j,k) :\  \mathrm{a\ permutation \ of\ (1,2,3)}.
\end{equation}
These relations, which are quite similar to Eqs.~(\ref{eqn:RestrictionAboutGamma}), hold for any Lindblad-type master equations for $2$-level systems. However, the inverse relaxation times ${\mit\Gamma}_i\ (i=1\sim3)$ are defined as the real parts of the eigenvalues $\mathrm{Re}\lambda_i \ (i=1,2,3)$ of the matrix $A$ in (\ref{eqn:MatrixBlochEq}): ${\mit\Gamma}_i \equiv \mathrm{Re} \lambda_i$ and the parameters $\gamma_i$'s are not the same as ${\mit\Gamma}_i$'s in general. Gorini {\it et~al.} have shown that if condition (\ref{eqn:Condition}) holds, $\gamma_i$'s coincide with ${\mit\Gamma}_i$'s, thus relations (\ref{eqn:RestrictionAboutGamma}) are derived directly from Eqs.~(\ref{eqn:RestrictionAboutgamma}) with $ {\mit\Gamma}_i= \gamma_i \ (i=1 \sim 3)$.

In what follows we generalize this result, i.e., we prove Eqs.~(\ref{eqn:RestrictionAboutGamma}) without resort to condition (\ref{eqn:Condition}). First the positivity of inverse relaxation times ${\mit\Gamma}_i \ge 0 \ (i=1,2,3)$ in Eqs.~(\ref{eqn:RestrictionAboutGamma}) must be satisfied, because a negative inverse relaxation time would result in a contradiction against the probability interpretation \cite{not:Positivity}. In order to prove the remaining inequalities in (\ref{eqn:RestrictionAboutGamma}):
\begin{equation}\label{eqn:RestrictionAboutGamma2}
{\mit\Gamma}_i + {\mit\Gamma}_j \ge {\mit\Gamma}_k, \quad (i,j,k) :\  \mathrm{a\ permutation \ of\ (1,2,3)},
\end{equation}
consider the following two cases separately: Case (i) There are one real eigenvalue $\lambda_{\mathrm{R}}$ and two complex eigenvalues $\lambda_{\mathrm{C}\pm} = \lambda_{\mathrm{CR}} \pm i\Omega$, conjugate to each other. Case (ii) There are three real eigenvalues $\lambda_{\mathrm{R}i} \ (i=1\sim 3)$. From the above discussion, we know that the real parts of the matrix $A$ in (\ref{eqn:MatrixBlochEq}) are positive: $\lambda_{\mathrm{R}}, \lambda_{\mathrm{CR}} \ge 0$ in Case (i) and $\lambda_{\mathrm{R}i} \ge 0 \ (i=1\sim 3)$ in Case (ii). In Case (i), two of Eqs.~(\ref{eqn:RestrictionAboutGamma2}) turns out to be equivalent to $\lambda_{\mathrm{R}}\ge 0$; non-trivial inequality is the remaining one:
\begin{equation}\label{eqn:2GR>GT}
2 \lambda_{\mathrm{CR}} \ge \lambda_{\mathrm{R}} \Leftrightarrow \tr A /2 \ge \lambda_{\mathrm{R}}.
\end{equation}
On the other hand, for any real $\lambda$,
\begin{equation}\label{eqn:Statement(i)}
\lambda \ge \lambda_{\mathrm{R}} \Leftrightarrow f(\lambda) \ge 0
\end{equation}
holds because $f(\lambda) = \det (\lambda - A) = 0$ has only one real solution $\lambda = \lambda_R$ and $\lim_{\lambda \to \pm \infty}f(\lambda) = \pm \infty$. This means that Eq.~(\ref{eqn:2GR>GT}) is equivalent to
\begin{equation}\label{eqn:ConditionForTheRG}
f(\tr A /2) \ge 0.
\end{equation}
Thus Eqs.~(\ref{eqn:RestrictionAboutGamma2}) are equivalent to Eq.~(\ref{eqn:ConditionForTheRG}), under the condition of positive inverse relaxation times $\lambda_{\mathrm{R}},\lambda_{\mathrm{CR}} \ge 0$ in Case (i). Quite similarly, in Case (ii), an equivalent condition to Eqs.~(\ref{eqn:RestrictionAboutGamma2}) is shown to be given by Eq.~(\ref{eqn:ConditionForTheRG}) provided all $\lambda_{{\mathrm R}i}$'s are positive \cite{ref:ProofFor(ii)1}. Then we have only to show (\ref{eqn:ConditionForTheRG}) for the matrix $A$ in (\ref{eqn:MatrixBlochEq}) to prove Eqs.~(\ref{eqn:RestrictionAboutGamma2}). A direct calculation shows
\begin{eqnarray}\label{f(trA/2)}
&&\quad f(\tr A /2) = \nonumber \\
&&\frac{1}{8}\Big[(\gamma_1 + \gamma_2 - \gamma_3)(\gamma_2 + \gamma_3 - \gamma_1)(\gamma_3 + \gamma_1 - \gamma_2)  \nonumber \\
&& \quad {} +  h^2_3(\gamma_1 + \gamma_2 - \gamma_3) + h^2_1(\gamma_2 + \gamma_3 - \gamma_1)  \nonumber \\
&&  \quad \quad \quad \quad \ \quad \quad \quad   \quad \quad{} +  h^2_2(\gamma_3 + \gamma_1 - \gamma_2)\Big].
\end{eqnarray}
From Eq.~(\ref{eqn:RestrictionAboutgamma}), which is a consequence of the completely positive condition, it is clear that this quantity is positive and the condition (\ref{eqn:ConditionForTheRG}) dose hold. This completes the proof of Eqs.~(\ref{eqn:RestrictionAboutGamma2}), and hence relations (\ref{eqn:RestrictionAboutGamma}) hold in any Lindblad-type master equations for $2$-level systems, irrespectively of condition (\ref{eqn:Condition})( See appendix A for a different proof of this).

We have shown that relations (\ref{eqn:RestrictionAboutGamma}) hold for any Lindblad-type master equations for $2$-level systems. Since the inverse relaxation times are physically meaningful and experimentally observable quantities \cite{ref:2GT>=GL}, the relations may serve as an experimental test ground to examine the validity of the Lindblad-type master equations. Furthermore, since the essential conditions for the validity of relations (\ref{eqn:RestrictionAboutGamma}) are Eqs.~(\ref{eqn:RestrictionAboutgamma}), which are direct consequence of the requirement of the completely positive condition, the relations may also give an experimental test \cite{ref:Benatti2} for the validity of the completely positive condition. Notice that a usage of completely positive condition, especially in a treatment of reduced dynamics with initial correlations \cite{ref:Royer}, is not fully justified \cite{ref:Pechukas}. Actually, it has shown \cite{ref:Buzvek} that the general form of a time evolution map with an initial correlation is not of the Kraus representation, which is the general form of the completely positive linear map with trace and positive preserving properties \cite{ref:Alicki}. Moreover, if there exists a non-unitary evolution that cannot be described in terms of reduced dynamics, then the use of a completely positive map for it might be illogical. In particular, in quantum measurement theory, the mechanism of decoherence still remains an open problem \cite{ref:Despagnat,ref:Zurek}. From these points of view, the role of such relations like (\ref{eqn:RestrictionAboutGamma}) is considered to be remarkable, for experiments may give us a convincing answer. Notice that it is well known that the relation (\ref{eqn:RestrictionAboutTLandTT}) is confirmed experimentally in all known cases \cite{ref:Alicki,ref:2GT>=GL}. We can say that, at least among these experiments, the results support the validity of the Lindblad-type master equation, and thus of the completely positive condition.

The author acknowledges useful comments and helpful discussions with Professors I. Ohba, S. Tasaki and M. Ohya. He is grateful to Professors H. Nakazato and A. Kossakowski for reading the manuscript prior publication and fruitful advice. He also thanks K. Imafuku, K. Yuasa, M. Miyamoto, Y. Ota, and B.~H. Valtan for useful comments and discussion.


\appendix

\section{}

In this appendix, we show more direct proof of relations (\ref{eqn:RestrictionAboutGamma}). We put $F_i = \sigma_i/2$ as before, but the positive matrix $[C_{ij}]$, written explicitly as
\begin{equation}\label{eqn:[C2]}
[C_{ij}] = \left(
 \begin{array}{ccc}
 \gamma_2+\gamma_3 -\gamma_1 & 2\delta_3 -ia_3 & 2\delta_2 + ia_2 \\
 2\delta_3 + ia_3 & \gamma_3 + \gamma_1 - \gamma_2 & 2\delta_1 -ia_1 \\
 2\delta_2 -ia_2 & 2\delta_1 + ia_1 & \gamma_1+ \gamma_2 -\gamma_3\end{array}\right),
\end{equation}
where $ a_i$'s, $\gamma_i$'s and $\delta_i$'s are real parameters, instead of (\ref{eqn:[C]}). Notice that the positive condition for matrix (\ref{eqn:[C2]}) requires $ C_{ii} \ge 0 \ (i=1 \sim 3)$ (the positivity of diagonal elements of $[C_{ij}]$ ), which follows (\ref{eqn:RestrictionAboutgamma}) as before, and also $ [\mathrm{adj} C]_{ii} \ge 0\  (i=1 \sim 3)$ (the positivity of diagonal elements of $\mathrm{adj} C$, which follows
\begin{equation}\label{eqn:RestrictionB_3}
\gamma^2_3 - (\gamma_1-\gamma_2)^2 - 4\delta^2_3 \ge a^2_3,
\end{equation}
especially for $i=3$ \cite{ref:Alicki}.

Expanding $H$ in terms of $F_i$, $H = \sum_{i=1}^3 h_i F_i$, with real parameters $h_i$'s, the generalized Bloch equation for the polarization components $M_i(t) = \tr \rho(t) F_i$ is shown to be
\begin{subequations}\label{eqn:MatrixBlochEq2}
\begin{equation}
\frac{d}{dt} \bm{M}(t) = - A^{\prime} \bm{M}(t) + \bm{b},
\end{equation}
\begin{equation}
A^{\prime}=\left(
\begin{array}{ccc}
\medskip
 \gamma_1 &  -\delta_3 + h_3 & -\delta_2 -h_2 \\
\medskip
- \delta_3 -h_3 & \gamma_2 & -\delta_1 + h_1 \\
-\delta_2 + h_2 & -\delta_1 -h_1 & \gamma_3
\end{array}
\right), \ {\bm b} = \frac{1}{2}\left(\begin{array}{ccc} a_1 \\ a_2 \\ a_3\end{array} \right).
\end{equation}
\end{subequations}
Since all the elements of matrix $A^{\prime} \in M(3)$ is real, there is at least one real eigenvalue whose eigenvector is also real. If we put a $z$-axis parallel to this eigenvector by transforming $F_i$ by some orthogonal matrix, $A^\prime$ can be rewritten as
\begin{equation}\label{NewA}
A^{\prime}=\left(
\begin{array}{ccc}
\medskip
 \gamma_1 &  -\delta_3 + h_3 & 0  \\
\medskip
- \delta_3 -h_3 & \gamma_2 & 0 \\
-2\delta_2 & -2\delta_1 & \gamma_3
\end{array}
\right),
\end{equation}
in this basis (with $-\delta_2-h_2 = -\delta_1 + h_1 = 0$). The eigenvalues of $A^{\prime}$ in (\ref{NewA}) are $\gamma_3$ and\begin{equation}
\frac{1}{2}(\gamma_1+\gamma_2) \pm \frac{1}{2}\sqrt{(\gamma_1-\gamma_2)^2 - 4(h^2_3 - \delta^2_3)}.
\end{equation}
We can prove relations (\ref{eqn:RestrictionAboutGamma}) by evaluating these eigenvalues directly, subject to the positivity of the matrix (\ref{eqn:[C2]}). Consider the following two cases separately: Case (a) $(\gamma_1-\gamma_2)^2 - 4(h^2_3 - \delta^2_3) < 0$ and Case (b) $(\gamma_1-\gamma_2)^2 - 4(h^2_3 - \delta^2_3) \ge 0$. For Case (a), we can put inverse relaxation times as ${\mit{\Gamma}}_1={\mit{\Gamma}}_2 = \frac{1}{2}(\gamma_1+\gamma_2)$ and ${\mit{\Gamma}}_3 = \gamma_3$. Then, the relation (\ref{eqn:RestrictionAboutGamma}) are equivalent to $\gamma_1+\gamma_2 \ge \gamma_3 \ge 0$, which holds from (\ref{eqn:RestrictionAboutgamma}). For Case (b), we can put inverse relaxation times as ${\mit{\Gamma}}_1= \frac{1}{2}(\gamma_1+\gamma_2) + \frac{1}{2}\sqrt{(\gamma_1-\gamma_2)^2 - 4(h^2_3 - \delta^2_3)}, \ {\mit{\Gamma}}_2 = \frac{1}{2}(\gamma_1+\gamma_2) - \frac{1}{2}\sqrt{(\gamma_1-\gamma_2)^2 - 4(h^2_3 - \delta^2_3)}$ and  ${\mit{\Gamma}}_3 = \gamma_3$. Similarly to Case (a), some of relations (\ref{eqn:RestrictionAboutGamma}): ${\mit{\Gamma}}_1 + {\mit{\Gamma}}_2 \ge {\mit{\Gamma}}_3 \ge 0, \ {\mit{\Gamma}}_1 \ge 0$ and ${\mit{\Gamma}}_3 + {\mit{\Gamma}}_1 \ge {\mit{\Gamma}}_2$ are shown to be hold from (\ref{eqn:RestrictionAboutgamma}). The rest inequalities ${\mit{\Gamma}}_2 + {\mit{\Gamma}}_3 \ge {\mit{\Gamma}}_1 \Leftrightarrow \gamma^2_3 - (\gamma_1-\gamma_2)^2 - 4\delta^2_3 + 4h^2_3 \ge 0$ and ${\mit{\Gamma}}_2 \ge 0 \Leftrightarrow \gamma_1\gamma_2 - \delta^2_3 + h^2_3 \ge 0$ also satisfy, since $\gamma^2_3 - (\gamma_1-\gamma_2)^2 - 4\delta^2_3 \ge 0$ and $\gamma_1\gamma_2 - \delta^2_3 \ge 0$ \cite{note:ProofOfThisIneq} hold from condition (\ref{eqn:RestrictionAboutgamma}) and (\ref{eqn:RestrictionB_3}). This completes the proof of relations (\ref{eqn:RestrictionAboutGamma}).



\begin{thebibliography}{10}
\bibitem{note:BH}
A map $\Phi$ on $B(\mathcal{H}_S)$ is called a completely positive map if the compound map $\Phi\otimes I_{n}$ on $B({\mathcal H}_S)\otimes M(n)$ preserves positivity for any $n \ge 1$. Here $\mathcal{H}_S$ denotes the Hilbert space associated with the system $S$ and $B(\mathcal{H}_S)$ the set of linear and bounded operators on $\mathcal{H}_S$, and $M(n)$ the algebra of $n\times n$ complex matrices with its identity map $I_{n}$.
\bibitem{not:FinDim}
Here we consider $N$-level systems. See Ref. \cite{ref:Lindblad} for the case of infinite-level systems.
\bibitem{note:BlochEq}
 Here $\Omega$ is a renormalized frequency and $\bm{b}$ a constant vector which determines the stationary state.
\bibitem{note:Normalization}
Here we use a different normalization, i.e., $\tr F_iF_j = 2\delta_{ij}$, as is used in \cite{ref:Gorini}.
\bibitem{note:Axis}
The matrix $[C_{ij}]$ is positive, and hence is also Hermitian. Although most general Hermitian matrix requires 9 parameters instead of 6, we can always put $F_i = \sigma_i/2$ such that the off-diagonal elements of $[C_{ij}]$ are purely imaginary, after transforming with some orthogonal matrix. Notice that this diagonalization has already been done in (\ref{eqn:[C]}), without losing any generalities to the discussion. See also Eq. (\ref{eqn:[C2]}) in appendix A in which this diagonalization has not been done.
\bibitem{note:PositivityOfgamma}
For example, since $\gamma_1+ \gamma_2 \ge \gamma_3$ and $\gamma_2 + \gamma_3 \ge \gamma_1$ hold from the positivity of the diagonal matrix elements of (\ref{eqn:[C]}), $\gamma_1 + \gamma_2 \ge \gamma_3 \ge \gamma_1 - \gamma_2 \Rightarrow \gamma_2 \ge 0$ follows. Similarly, all $\gamma$'s are shown to be positive.
\bibitem{not:Positivity}

To show this more directly, the following proposition can be used: Let $A$ be a $3\times 3$ real matrix, and $\lambda_i\ (i=1,2,3)$ the eigenvalues of $A$. It is easy to verify that the necessary and sufficient condition for all of the real parts of the eigenvalues $\mathrm{Re} \lambda_i \ (i=1,2,3 )$ to be positive, is that $\tr A\ge0,\ \det A\ge0,\ \tr \mathrm{adj} A \ge0 $, and $f(\tr A)\ge0$, where $\mathrm{adj} A$ denotes the cofactor matrix of $A$ and $f(\lambda)=\det(\lambda-A)$. For the matrix $A$ in (\ref{eqn:MatrixBlochEq}), the direct calculation shows $ \tr A = \sum_{i=1}^{3}\gamma_i, \det A = \gamma_1 \gamma_2 \gamma_3 + \sum_{i=1}^{3} \gamma_i h^2_i$, $ \tr \mathrm{adj} A = (\gamma_1 \gamma_2 + \gamma_2 \gamma_3 + \gamma_3 \gamma_1) + \sum_{i=1}^{3}h^2_i$, $ f(\tr A) = \gamma_1\gamma_2(\gamma_1+\gamma_2) + \gamma_2\gamma_3(\gamma_2+\gamma_3) + \gamma_3\gamma_1(\gamma_3+\gamma_1) + \gamma_1(h^2_2+h^2_3)+ \gamma_2(h^2_3+h^2_1)+\gamma_3(h^2_1+h^2_2)+2\gamma_1\gamma_2\gamma_3$. Clearly they are all positive, since $\gamma_i \ge 0\ (i=1,2,3)$ from (\ref{eqn:RestrictionAboutgamma}) and  $h_i$'s are real. Then the positivity of the inverse relaxation times ${\mit\Gamma}_i = \mathrm{Re} \lambda_i \ge 0 \ (i=1,2,3)$ follows from the above proposition.
\bibitem{ref:ProofFor(ii)1}
In Case (ii), the non-trivial one of Eqs.~(\ref{eqn:RestrictionAboutGamma2}) is $\lambda_{\mathrm{R}2} + \lambda_{\mathrm{R}3} \ge \lambda_{\mathrm{R}1} \Leftrightarrow \tr A/2 \ge \lambda_{\mathrm{R}1} $ with denoting descending order: $\lambda_{\mathrm{R}1} \ge \lambda_{\mathrm{R}2} \ge \lambda_{\mathrm{R}3} \ge 0$. In this case, we cannot use the statement (\ref{eqn:Statement(i)}). Instead, we know that $\lambda \ge \lambda_{\mathrm{R}1}$ or $\lambda_{\mathrm{R}2} \ge \lambda \ge \lambda_{\mathrm{R}3} \Leftrightarrow f(\lambda) \ge 0$ holds for any real $\lambda$. But at a particular point: $\lambda=\tr A/2$, one can easily check that $\tr A/2 \ge \lambda_{\mathrm{R}1} \Leftrightarrow f(\tr A/2) \ge 0$, since if we assume that $f(\tr A/2) \ge 0$ and $ \lambda_{\mathrm{R}2} \ge \tr A/2 \ge \lambda_{\mathrm{R}3}$ hold, it does not contradicts only when $\tr A/2=\lambda_{\mathrm{R}2}=\lambda_{\mathrm{R}1}$, which can be included in $\tr A/2 \ge \lambda_{\mathrm{R}1}$.
\bibitem{note:ProofOfThisIneq}
(\ref{eqn:RestrictionB_3}) is equivalent to $4(\gamma_1\gamma_2 - \delta^2_3) \ge (\gamma_1 + \gamma_2)^2 - \gamma^2_3 + a^2_3$, and the right hand side of this inequality is positive from (\ref{eqn:RestrictionAboutgamma}).
\end{thebibliography}
\end{document}